\documentclass[11pt]{article}
\usepackage[T1]{fontenc}
\usepackage[utf8]{inputenc}
\usepackage{amsmath}
\usepackage[numbers]{natbib}
\usepackage{authblk}
\usepackage{hyperref}

\usepackage[english]{babel}
\usepackage[utf8]{inputenc}
\usepackage{amsmath}
\usepackage{xcolor}
\usepackage{graphicx}
\usepackage[colorinlistoftodos]{todonotes}
\usepackage{csquotes}
\usepackage{listings}
\usepackage{cleveref}
\usepackage{numprint}
\usepackage{authblk}

\crefname{appsec}{Appendix}{Appendices}

\newcommand{\M}[0]{\mathbf{M}}
\newcommand{\mb}[1] {\mathbf{#1}}

\newcommand{\field} {\mb{M} (\mb{r'})}
\newcommand{\R} {\mb{r}}
\newcommand{\Rs} {\mb{r}_s}

\newcommand{\Rrs} {|\R_t-\Rs|}

\newcommand{\grad} {\mb{\nabla}}
\newcommand{\kernel}[1][] {D#1_\mb{n}\left(\R\right)}
\newcommand{\kerneltwo}[2][] {D#1_\mb{n}\left(#2\right)}

\newcommand{\potKern} {\grad_{\!s}\frac{1}{\Rrs}}
\newcommand{\dd} {\mathrm{d}}
\newcommand{\potential}[1][] {u (\mb{r}#1)}

\newcommand{\BigO}[1]{\ensuremath{\mathcal{O}\bigl(#1\bigr)}}

\newcommand{\stdgrafx}[1]{\includegraphics[width=0.9\textwidth]{#1}}

\begin{document}
\bibliographystyle{plainnat}
\date{\today}

\title{Highly Parallel Demagnetization Field Calculation Using the Fast
  Multipole Method on Tetrahedral Meshes with Continuous Sources}

\author[1]{Pietro Palmesi}
\author[2,3]{Lukas Exl}
\author[1]{Florian Bruckner}
\author[1]{Claas Abert}
\author[1]{Dieter Suess}

\affil[1] {Christian Doppler Laboratory for Advanced Magnetic Sensing and Materials, Institute of Solid State Physics, TU Wien, Vienna, Austria}
\affil[2]{Institute of Solid State Physics, TU Wien, Vienna, Austria}
\affil[3]{Faculty of Mathematics, University of Vienna, Vienna, Austria}

\maketitle

\begin{abstract}The long-range magnetic field is the most time-consuming part in
micromagnetic simulations. Improvements both on a numerical and computational
basis can relief problems related to this bottleneck. This work presents an
efficient implementation of the Fast Multipole Method [FMM] for the magnetic
scalar potential as used in micromagnetics. We assume linearly magnetized
tetrahedral sources, treat the near field directly and use analytical
integration on the multipole expansion in the far field. This approach tackles
important issues like the vectorial and continuous nature of the magnetic field.
By using FMM the calculations scale linearly in time and memory. 
\end{abstract}

\section{Introduction}
Micromagnetic algorithms are an important tool for the simulation of
ferromagnetic materials used in electric motors, storage systems and magnetic
sensors. In micromagnetic simulations the demagnetization field is the most
time-consuming part. Many different algorithms for solving the problem exist, 
e.g.~\cite{fredkin_hybrid_1990, brunotte_finite_1992, buchau_comparison_2003,
popovic_applications_2004, beatson_short_1997, knittel_compression_2009,
blue_using_1991, livshitz_nonuniform_2009, abert_fast_2012, exl_fast_2012}.
Direct solutions compute all pair-wise interactions and hence they scale with
$\BigO{N^2}$ operations, where $N$ denotes the number of
interaction partners.

Modern Finite Element Method (FEM) implementations\cite{chang_fastmag_2011,
abert_fast_2012} have linear scaling for the Poisson equation
only for closed boundary conditions when a multigrid preconditioner is
used\cite{tsukerman_multigrid_1998}. The demagnetization potential has open
boundary conditions defined at infinity. For the computation of the potential at the boundary, hybrid FEM/BEM\cite{fredkin_hybrid_1990} (Finite Element Method with Boundary Element Method Coupling) and the shell
transformation approach \cite{brunotte_finite_1992} add an additional complexity of \BigO{M^2}, where $M$ denotes the number of surface triangles. This can be further reduced to \BigO{M \log(M)} by the use of
$\mathcal{H}$-matrix approximation \cite{buchau_comparison_2003,
popovic_applications_2004, knittel_compression_2009} or fast summation
techniques utilizing non-uniform fast Fourier transform (NUFFT)
\cite{exl_non-uniform_2014}.

The Fast Multipole Method (FMM) was originally invented by
\citeauthor{greengard_fast_1987}\cite{greengard_fast_1987} for particle
simulations. The version presented in this paper is an extension to the
class of Fast Multipole Codes~\cite{cheng_fast_1999,
  raykar_short_2005, yokota_fmm_2013}. This version is based on the direct
evaluation of volume integrals, putting it in a category with nonuniform grid
(\cite[]{livshitz_nonuniform_2009}), Fast Fourier Transform [FFT] methods
(\cite{long_fast_2006, abert_fast_2012}) and Tensor Grid methods (\cite{exl_fast_2012}).
A useful comparison of existing codes is given in the review~\cite{abert_numerical_2013}. 
The basis of the presented approach is a spatial discretization of the analytical expression of the integral representation of the magnetic scalar potential
\begin{align}
 \potential = \frac{1}{4\pi}\int_\Omega \field \cdot \nabla^\prime\Big(\frac{1}{|\mb{r} - \mb{r}^\prime|} \Big)\,\text{d}\mb{r}^\prime \\
\approx \frac{1}{4\pi}\sum_{j=1}^{M}\int_{\Omega_j} \mb{M}_j(\mb{r}^\prime) \cdot \nabla^\prime\Big(\frac{1}{|\mb{r} - \mb{r}^\prime|} \Big)\,\text{d}\mb{r}^\prime,\label{eq:approx1}
\end{align}
where the $M$ subvolumes $\Omega_j$ are tetrahedrons wherein the magnetization
$\mb{M}_j$ is assumed to be linear (i.e.affine basis functions). The FMM
approach in this work approximates~\eqref{eq:approx1} by treating the near field
both directly and analytically and the far field by (exact) numerical integration of a
multipole expansion of the integrated convolution kernels.

Among the above mentioned methods only the FEM/BEM approach and a non-uniform
FFT\cite{exl_non-uniform_2014} approach are suited for irregular grids, but FMM
is expected to scale better to many processors because of its small memory
footprint. Some implementations of current micromagnetic FMM
codes~\cite{zhang_adaptation_2009, blue_using_1991} only support homogeneously
magnetized regular grids. The hereby presented method works on linearly
magnetized tetrahedrons, making it a suitable alternative to FEM codes on
irregular grids. However, the tetrahedral nature of the sources complicates the
FMM procedure by demanding more elaborate approaches for the direct interaction
(see \cref{sec:direct}) and multipole expansion (see \cref{sec:expansion}). This
work manages to preserve the optimal linear scaling of FMM codes with good
performance even for small problems (see \cref{fig:solvetime}) while using low
storage compared to most other algorithms. The storage requirements are
particularly important for large problems as well.

The paper is divided as follows. We present the relevant mathematical and
algorithmic preliminaries of the method in the next section. Implementation
details are given in \cref{sec:implementation}. In the numerics section
(\cref{sec:numerics}) we compare the results for the problem of a homogeneously
magnetized box to the analytic solution.
Details related to analytical integration over tetrahedrons
are stated in \cref{app:directIntegrals}.

\section{Fast Multipole Expansion [FMM]}\label{fast-multipole-expansion-fmm}

\begin{figure}
\stdgrafx{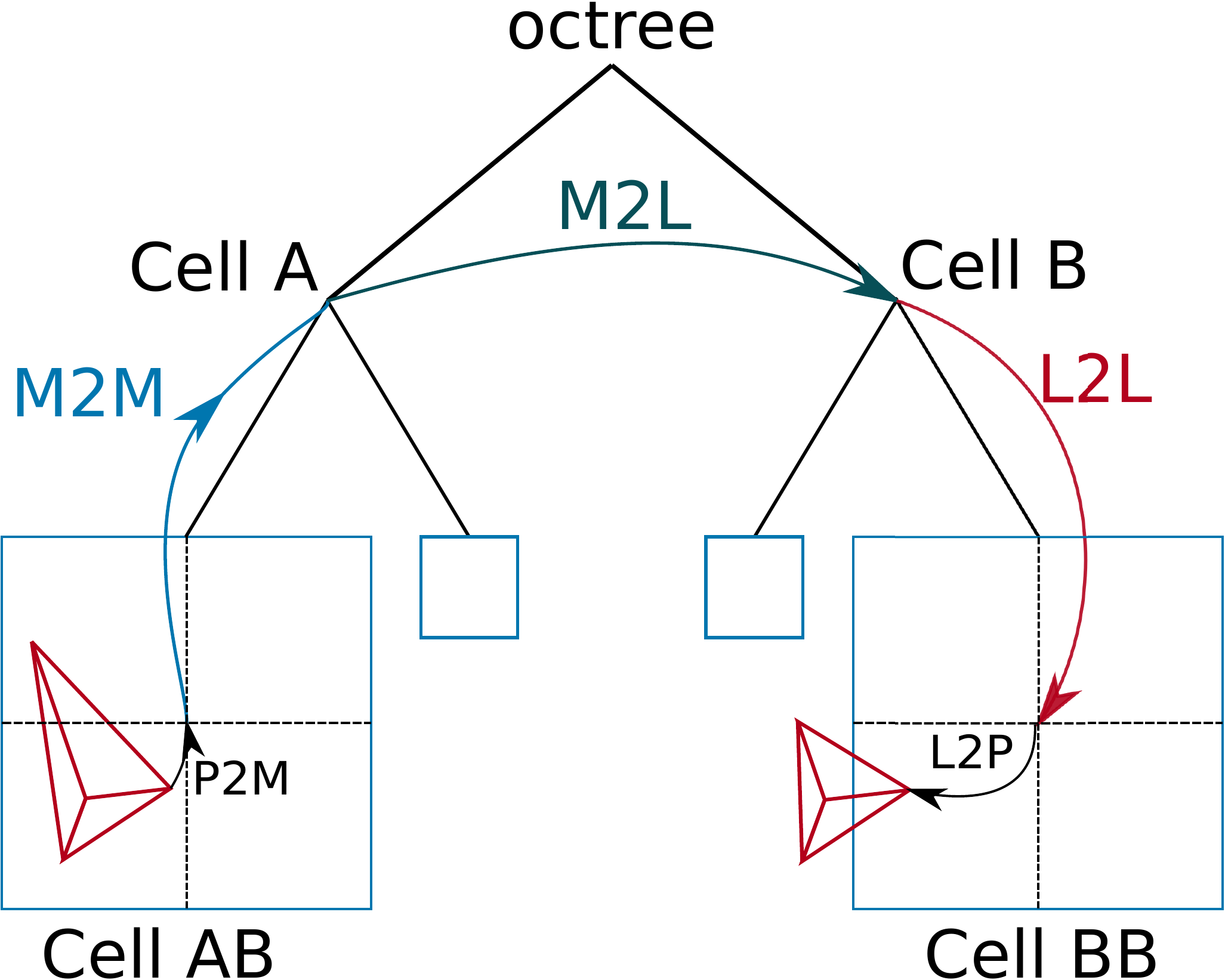}
\caption{Overview of the FMM algorithm showing the involved operations and multipole transitions. P~\dots~Point~(vertex,~source~or~target), M~\dots~Multipole~expansion~of~source, L~\dots~Local~expansion~around~target~cell}
\label{fig:traversal}
\end{figure}

The fast multipole method is a method for computing equations of the
form 
\[ u_i =\sum_j \kerneltwo[]{s_j-t_i} Q_j, \]
where $u_i$ is the potential at the $i$th evaluation position, \(\kernel\) is a so called kernel function, \(t_i\)
are called target-, \(s_j\) source-points and \(Q_j\) magnetizations or charges. For a
dense kernel, this computation would need \(\BigO{ N^2 }\) operations, compared
to the \(\BigO{ N }\) for the FMM. As pointed out in~\cite{beatson_short_1997}, the key features of the
FMM are:

\begin{itemize}
\itemsep1pt\parskip0pt\parsep0pt
\item
	A specified acceptable accuracy of computation \(\epsilon\) (in this case a flexible
  Multipole Acceptance Criterion for adjusting speed vs.\ accuracy; see~\cref{sec:mac}~and~\cite{yokota_fmm_2013})
\item
	A hierarchical division of space into source clusters (octree)
\item
	A multipole expansion of the kernel \(\kernel\)
\item
  (for improved performance) A method for calculating the local
  expansion from the far-field expansion~\cite{visscher_simple_2010} 
\end{itemize}
\noindent
A sketch of the the FMM algorithm is shown in~\cref{fig:traversal}. It
involves either the direct computation of the interaction of two vertices
(P2P~\dots~Point to Point), or the following 5 steps:
\begin{enumerate}
	\item Multipole expansion of the source vertex into the leave cell (P2M~\dots~Point to Multipole)
	\item Collection of smaller cell multipole moments into combined multipole
    moments of the parent cell (M2M~\dots~Multipole to Multipole)
	\item Local expansion of remote multipole moments (M2L~\dots~Multipole to Local)
	\item Down conversion of local expansion coefficients to the descendant cells. (L2L~\dots~Local to Local)
	\item Evaluation of the local expansion at the target vertex. (L2P~\dots~Local to Point)
\end{enumerate}
where steps 2 and 4 can be done as many times as necessary.

\subsection{Direct Vertex Interaction}\label{sec:direct}
Each tetrahedron in the mesh is split into four tetrahedral hat functions. These
hat functions are then identified with their vertices. The integration from one
vertex $v_1$ to another vertex $v_2$ is then carried out by integrating over all
hat functions that are non-zero at $v_1$. This leads to the integration shown
in~\cref{eq:approx1}. The integral (solved in the sections below) results in a
vector that describes the interaction from one vertex to another. For a similar
approach consult \cite*{graglia_static_1987}. In this method, targets are
treated as simple vertices. A recently published method for integrating the
interaction between two polyhedrons exists\cite{chernyshenko_computation_2016}.

\subsubsection{Tetrahedron Integration Overview}
\begin{figure}
\stdgrafx{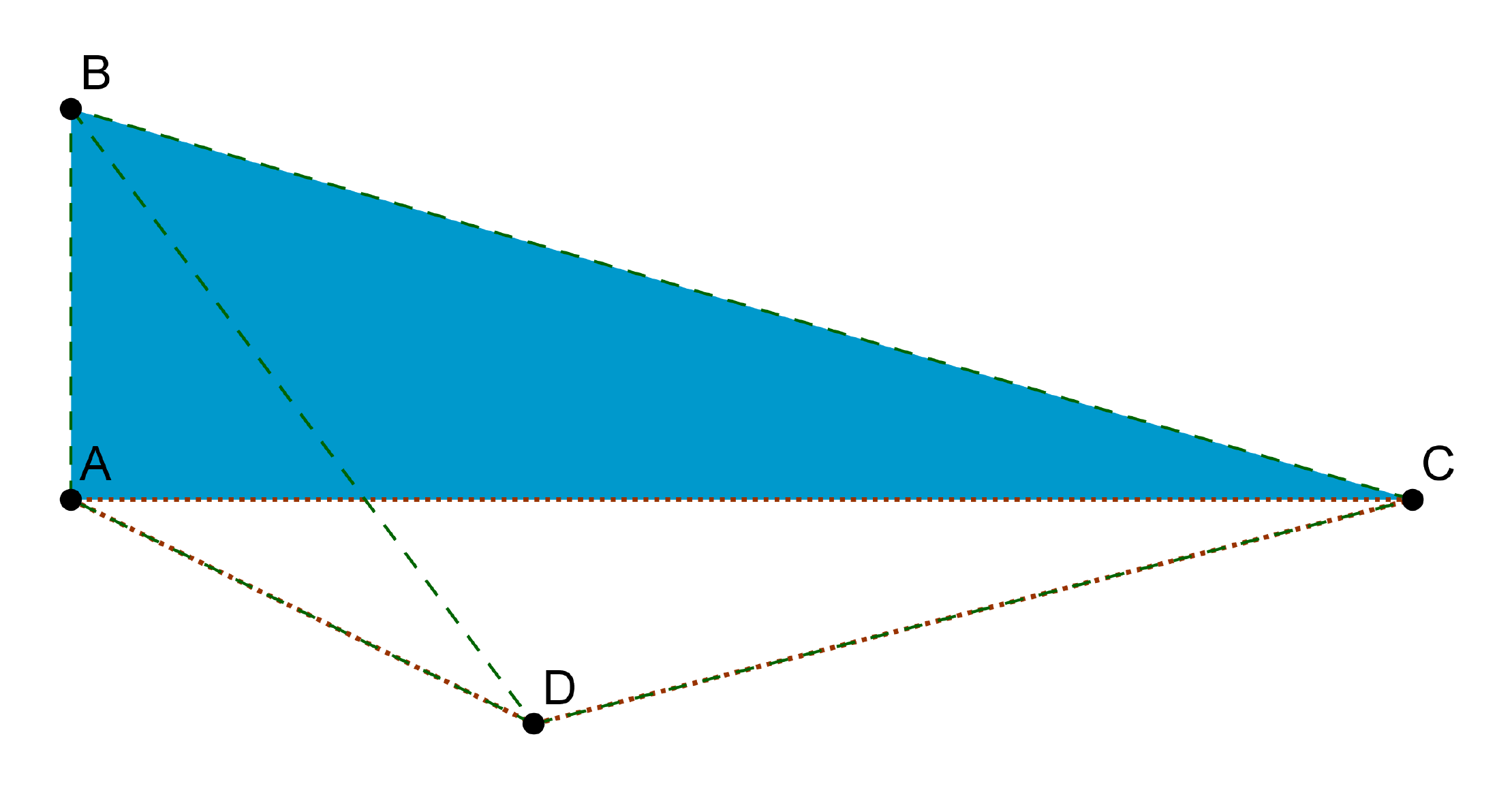}
\caption{Decomposition of the source triangle integration (ABC;\;blue triangle) into
  three new triangles with the target (D) as vertex. The integrals described by the triangles (ABD, BCD) are
  added and the integral bounded by triangle (ACD) is subtracted.}
\label{fig:decomposition}
\end{figure}

\begin{figure}
\includegraphics[width=0.5\textwidth]{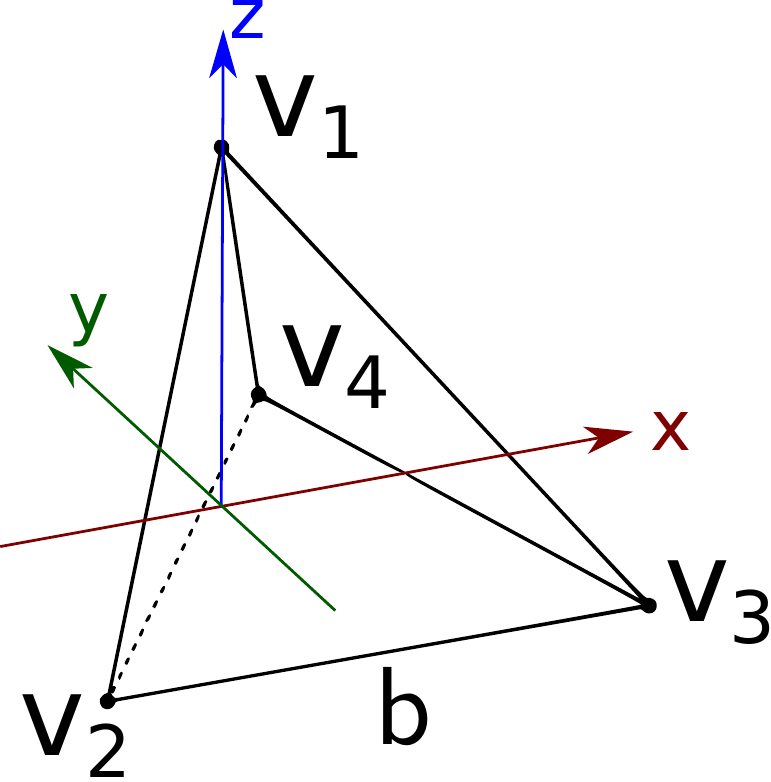}
\caption{The rototranslated frame of reference for tetrahedron integration. With
$x_1 = y_1 = 0$, $z_2=z_3=z_4=0$ and $y_2=y_3$}
\label{fig:coords}
\end{figure}

Integration of the kernel function over any tetrahedral domain can be decomposed
to four integrals with the target point replacing one source vertex for each
vertex\cite{graglia_static_1987}. A two-dimensional integral decomposition
can be seen in \cref{fig:decomposition}. This step simplifies the problem into
four integrations of tetrahedrons with an $\R$ at a corner vertex.
Before integrating the tetrahedrons the problem is rototranslated into a new frame
of reference sketched in \cref{fig:coords}. The target point $v_1$ is set
into the z-axis. The other vertices are set into the xy-plane and the source
vertices $v_2$ and $v_3$ are aligned parallel to the x-axis.

\subsubsection{Source Magnetization}
The integrals can be solved for a general linear magnetization. To this end, 
the magnetization $\field$ is split into a geometric part describing
the linear change of the field and a value part describing the magnetization at
each vertex. The hat functions set all but one source vertex to zero.
Giving at most two nonzero magnetized vertices, namely the source vertex $v_1$ and the
target vertex $v_4$. Without loss of generality the system of linear equations for
the x-component of the field is:
$$\M_{\hat{x}}(x,y,z) = a_1 + a_2x + a_3y + a_4z = a_{\hat{x}} \cdot {(1,x,y,z)}^T$$
$$\begin{aligned}
\M_x\left(v_1=(0 ,0 ,h=1)\right) &=M_t \\
\M_x\left(v_2=(x_2,y_2,0)\right) &=0 \\
\M_x\left(v_3=(x_2,y_3,0)\right) &=0 \\ 
\M_x\left(v_4=(x_4,y_4,0)\right) &=M_s 
\end{aligned}$$
where $M_t$ is the linear source field evaluated at the target vertex. And $M_s$ is
the magnetization of the source vertex, which can be expressed in matrix form as
$$
\begin{pmatrix}
1 & 0 & 0 & 1 \\
1 & x_2 & y_2 & 0 \\
1 & x_2 & y_3 & 0 \\
1 & x_4 & y_4 & 0
\end{pmatrix}
\cdot
\begin{pmatrix}
a_1 \\ a_2 \\ a_3 \\ a_4
\end{pmatrix}
= \begin{pmatrix}
M_t \\ 0 \\ 0 \\ M_s
\end{pmatrix}
$$
solving to:
$$a_{\hat{x}}= \frac{1}{bs} \begin{pmatrix}
M_s (x_2-x_3) y_2 \\
0 \\
M_s (x_3-x_2) \\
\left( M_s (x_3-x_2) + M_t (x_3-x_2) (y_4-y_2) \right)
\end{pmatrix}$$ with $b = x_3 - x_2$ and $s = y_4 - y_2$

\subsubsection{Tetrahedron Corner Integration}
The rototranslated tetrahedron integrals are all singular but can be transformed into
regular integrals by following nonlinear transformation (see also
\cite{graglia_static_1987}):
$$x\rightarrow x_t (1-\lambda); y\rightarrow y_t (1-\lambda); \frac{z}{h}\rightarrow \lambda$$
The transformed integral creates a prism which can be integrated in $\lambda$-direction
leading to following triangle integration:
\begin{equation}
  \begin{aligned}
\potential &= \frac{1}{4\pi}\int_\Omega \field \cdot \potKern \dd \mb{r}' \\
 &= \frac{1}{4\pi} \mb{a}\int_T\frac{x_t \hat{x} + y_t \hat{y} +
  h\hat{z}}{{\left( x_t^2+y_t^2+h^2 \right)}^{\frac{3}{2}}}
  \begin{pmatrix} 
  1 \\
  x_t=0 \\
  y_t \\
  1
\end{pmatrix}
\dd x_t \dd y_t
  \end{aligned}
\label{eqn:tri_potential}
\end{equation}
Note that $\mb{a}$ is a $3 \times 4$ matrix contracting with $\hat{x},\hat{y},
\hat{z}$ and ${(1,0,y_t,1)}^T$.

The triangle integration boundaries have been set to:
\begin{itemize}
\item x-integration first ($\int_T\dd x_t \dd y_t= \int_{y-}^{y+}\int_{x-}^{x+}
  \dd x_t \dd y_t$):
$$x_{-}=\frac{x_1-x_2}{y_1-y_2}(y-y_2)+x_2;\; x_{+}=\frac{x_1-x_3}{y_1-y_2}(y-y_2)+x_3$$
$$y_{-}=y_2;\; y_{+}=y_3$$
\item y-integration first ($\int_T \dd y_t\dd x_t=
  \int_{x-1}^{x+1}\int_{y-}^{y+}\dd y_t\dd x_t+\int_{x-2}^{x+2}\int_{y-}^{y+}\dd
  y_t\dd x_t$):
$$y_{y-}=\frac{y_1-y_2}{x_1-x_2}(x-x_2)+y_2;\; y_{y+}=\frac{y_1-y_2}{x_1-x_3}(x-x_1)+y_1$$
$$x_{y-1}=x_1;\; x_{y+1}=x_2$$
$$x_{y-2}=x_2;\; x_{y+2}=x_3$$
\end{itemize}

The resulting triangle integrals from \cref{eqn:tri_potential} are given in
\cref{app:directIntegrals}.

\subsection{Source Expansion}\label{sec:expansion}
For continuous sources an integral expression (in our case \cref{eq:approx1}) must
be expanded. To this end the Coulomb kernel $1/|\mb{r}-\mb{r}^\prime|$ is
expanded as a Taylor polynomial of order $p$ in terms of $\mb{r}^\prime$ around
the origin (\textit{Cartesian multipole expansion})
\begin{align} \frac{1}{|\mb{r} - \mb{r}^\prime|} = \sum_{|\mb{n}|\leq p}
\frac{1}{\mb{n}!}\,\Big(\frac{\partial^{\mb{n}}}{\partial{(-\mb{r})}^{\mb{n}}}\,
\frac{1}{|\mb{r}|}\Big){\mb{r}^\prime}^{\mb{n}} +
\mathcal{O}(\tfrac{|\mb{r}^\prime|^{p+1}}{|\mb{r}|^{p+2}}).
\end{align} where $\mb{n} = {n_x,n_y,n_z}$ being a multi-index,
$\mb{n}!=n_x!n_y!n_z! $, $|\mb{n}| = n_x+n_y+n_z$ and
$\R'^{\mb{n}}=x'^{n_x}y'^{n_y}z'^{n_z}$. From a source point (or field point,
resp.) perspective the truncation error is proportional to ${(\alpha/2)}^{p+1}$
where $\alpha$ is the opening angle $2r/d$ with $r$ being the diameter of source
(or field) cell and $d$ the distance between centers
\cite{visscher_simple_2010}. Plugging the expansion into the expression for the
$j$-th volume source (\cref{eq:approx1}) yields
\begin{align}
 \int_{\Omega_j} \mb{M}_j(\mb{r}^\prime) \cdot \nabla^\prime \Big(
\frac{1}{|\mb{r} - \mb{r}^\prime|}\Big)\,\text{d}\mb{r}^\prime =
\sum_{|\mb{n}|\leq p} \frac{1}{\mb{n}!} \Big(
\frac{\partial^{\mb{n}}}{\partial{(-\mb{r})}^{\mb{n}}}
\frac{1}{|\mb{r}|}\Big)\int_{\Omega_j} \mb{M}_j(\mb{r}^\prime) \cdot
\nabla^\prime \R'^{\mb{n}}\,\text{d}\mb{r}^\prime,
\end{align}
resulting in the following expansion coefficients:
\begin{equation}
Q_{\mb{n};j} = \int_{\Omega_j}  \mb{M}_j(\mb{r}^\prime) \cdot \nabla^\prime \R'^{\mb{n}}\,\text{d}\mb{r}^\prime.
\label{eqn:expCoeffs}
\end{equation}


The expansion coefficients can be calculated \textit{exactly} using a quadrature rule of
order $M=|\mb{n}|+1$ because of the polynomial nature of the integrand, i.e.
$$\int_{\Omega_j}f(r)\dd^3r=|\Omega_j|\sum_{i=1}^Mf(p_i)w_i,$$
using the volume of the tetrahedron $|\Omega_j|$ and the quadrature points $p_i$ with associated weights $w_i$ from~\cite{zhang_set_2009}.

\subsection{Local Expansion}
In the target octree-cell the potential $u(\R)$ is approximated by a power
series $$u(\R)=\sum_\mb{n}\frac{1}{\mb{n}!}L_{\Omega,n}r^n$$ again using $\mb{n}$ as a
multi-index (see~\cref{sec:expansion}).  This gives the local expansion coefficients
of $$L_{\Omega,n}=\left( \frac{\partial^{\mb{n}}}{\partial \R^{\mb{n}}} u(\R)
\right)\bigg|_{\R=r_{\Omega}}.$$
The local expansion coefficients $L_{\Omega,n}$ can be computed recursively (see~\cite{visscher_simple_2010}).

\section{Implementation}\label{sec:implementation}
\subsection{Hierarchical Space Division}

\begin{figure}
\begin{minipage}[b]{0.45\linewidth}
\stdgrafx{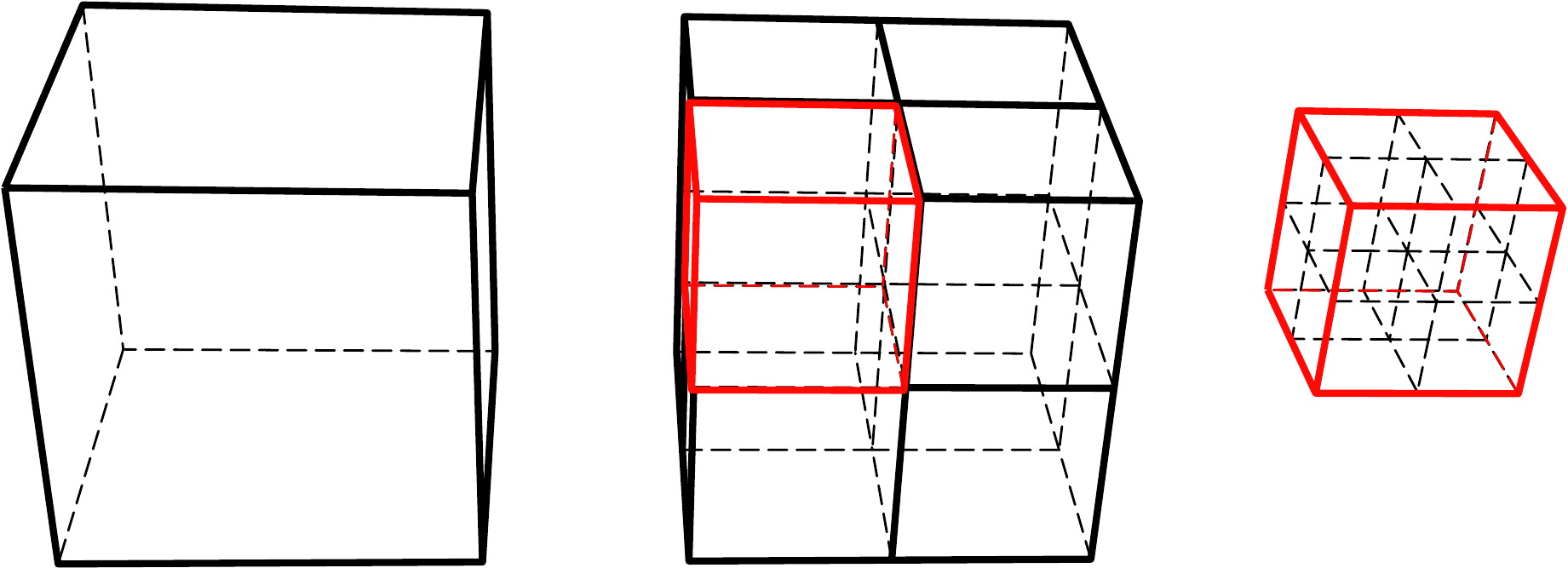}  
\caption{An octree is the successive subdivision of space into 8 smaller cubes.}
\label{fig:octree}
\end{minipage}
\hspace{0.5cm}
\begin{minipage}[b]{0.45\linewidth}
\stdgrafx{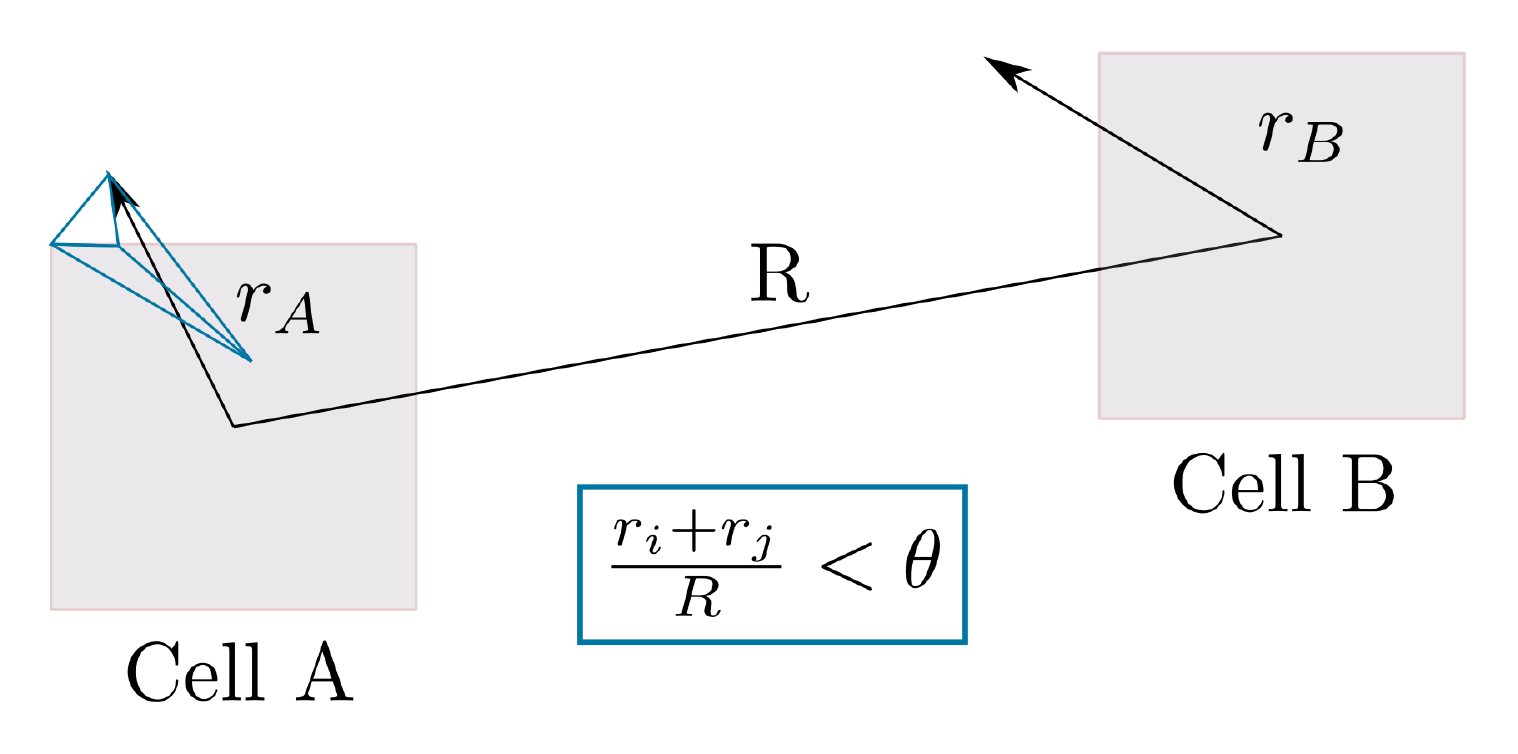}
\caption{Multipole Acceptance Criterion: $r_A$ and $r_B$ are the radii of the
remotest directly connected vertices, $R$ is the cell distance and $\theta$ the
MAC~parameter.}
\label{fig:mac}
\end{minipage}
\end{figure}

The key to the good scalability of the algorithm is the hierarchical space
division. In three dimensions a cubic box enclosing the investigated space is
successively divided into octrees creating smaller and smaller cubes of space
(see~\cref{fig:octree}). The smallest (leaf) cells contain the vertex points
of the mesh. These vertices are in turn connected to linear hat functions
defined by the adjacent tetrahedrons creating the irregular mesh.

\subsection{Multipole Acceptance Criterion [MAC]}\label{sec:mac}

The MAC is used to determine whether direct integration or multipole expansion is
used for two interaction partners\footnote{interaction partners can be either a
vertex or an octree-cell}. The MAC is fulfilled when the combined radii are
smaller than the interaction partners' distance by a factor of $\theta$
(see~\cref{fig:mac}). In this case expansion is used, otherwise direct calculation. The cell radius is the distance between the cell center
and the most remote vertex directly connected to the cell. A vertex is
directly connected if it shares a tetrahedron with a vertex residing inside
the cell. It is not the size of the cell itself.

\subsection{Dual Tree Traversal}
The dual tree traversal moves down the source and target tree simultaneously
using the MAC to decide whether to do a multipole to local expansion (M2L) or
move down a level in the larger tree and repeat the process for the newly
created cell pairs. When both trees are at the leaf level a direct computation
of the potential is done. The method and pseudocode for the dual tree traversal is reprinted
from \cite{yokota_fmm_2013} in~\cref{lst:dtt} and~\cref{lst:interact}.

\vbox{%
\lstset{caption={EvaluateDualTreecode{()}}, language=Pascal, label=lst:dtt}
\begin{lstlisting}
push pair of root cells (A,B) to stack
while stack is not empty do
  pop stack to get (A,B)
  if target cell is larger then source cell then
    for all children a of target cell A do
      Interact{(a,B)}
    end for
  else
    for all children b of source cell B do
      Interact{(A,b)}
    end for
  end if
end while
\end{lstlisting}}

\vbox{%
\lstset{caption={Interact{(A,B)}}, language=Pascal, label=lst:interact}
\begin{lstlisting}
if A and B are both leafs then
  call P2P kernel
else
  if A and B satisfy MAC then
    call M2L kernel
  else
    push pair (A,B) to stack
  end if
end if
\end{lstlisting}}
\subsection{Storage Requirements}
The algorithmic storage required is mainly composed of tree information, direct interaction
coefficients, multipole coefficients, sources and targets. Obviously the source
and target information scales linearly with the number of sources and targets. The
direct integration coefficients stored are mainly a near field component; they
scale linearly. The number of multipole and local coefficients also scales linearly
but the proof is a bit more nuanced. Consider a tree with N vertices. The number
of octree levels is equal to $\log_8(N/a)$ where $a$ is the number of vertices per
leaf cell. The number of nodes per level $L$ is $8^L$ giving a total of 
$$\sum_{n=1}^{\log_8(N)}8^n = \frac{1-8^{\log_8(N/8)+1}}{1-8}=\frac{N-1}{7}=O(N)$$
levels. Each level must store $p(p+1)(p+2)/6$ local and multipole coefficients. Which
means it scales cubic in the multipole order.
Summarizing, it can be said that the storage scales linearly in the number of
sources and targets, but with cubic dependence on the multipole order $p$.

\section{Numerics}\label{sec:numerics}
This section compares the computational results of \cref{eq:approx1} in an
applied physical context. The near-field (P2P) evaluation described
in~\cref{sec:direct} was verified with randomly magnetized and randomly generated
tetrahedra via numeric integration, leading to an error in the order of the
numeric accuracy of the quadrature.

\subsection{Comparison with FEM solvers}\label{sec:femComparison}
The present work implemented two implementations of the FMM for the use on
tetrahedral grids. One for linearly magnetized meshes (Tet~FMM) and another for point-like
dipole clouds (Mag~FMM). Focus was placed on speed and scaling. It shows that FMM is a
fast usable algorithm for solving the stray-field problem which can be
integrated in existing micromagnetic codes. Applications for the
point-like source FMM include atomistic simulations on many cores.

\begin{figure}
\stdgrafx{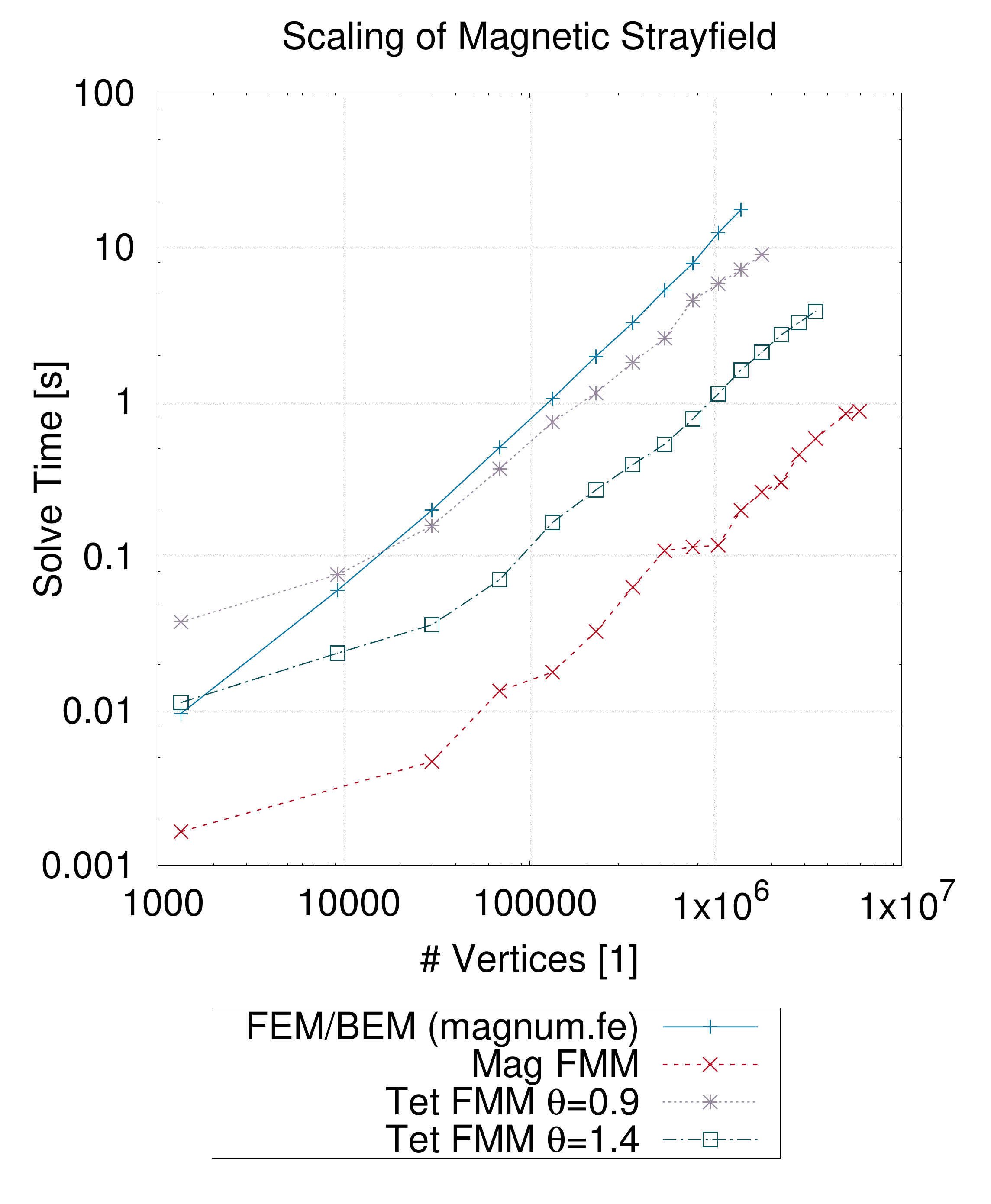}
\caption{This plot compares the runtime of a fast FEM code (magnum.fe) to the
run-time of a simple dipole FMM (Mag FMM) code and the improved FMM (Tet FMM)
code with direct near-field interaction and a MAC of 0.3 (see~\cref{sec:mac}).
Each simulation was done on one node with 16 cores.}
\label{fig:solvetime}
\end{figure}

Simulations on homogeneously magnetized regular boxes have been done on a
modified version of the ExaFMM\cite{yokota_fmm_2013}\footnote{Source:
https://github.com/exafmm/exafmm.git} software to determine accuracy and speed
of the produced code. The speed of these simulations was compared to an
optimized Finite Element Method [FEM] simulation\footnote{magnum.fe:
http://micromagnetics.org/magnum.fe/} in \cref{fig:solvetime}.

The parameters used for the Tet FMM code were a MAC of $\theta = 0.8$ and
$\theta=1.4$ and a maximum multipole order of $O=4$. The parameters used for the
Mag FMM code were $\theta=0.3$ and $O=4$. Mass lumping was used for the
computation of the B-field and the energy.

\begin{figure}
\stdgrafx{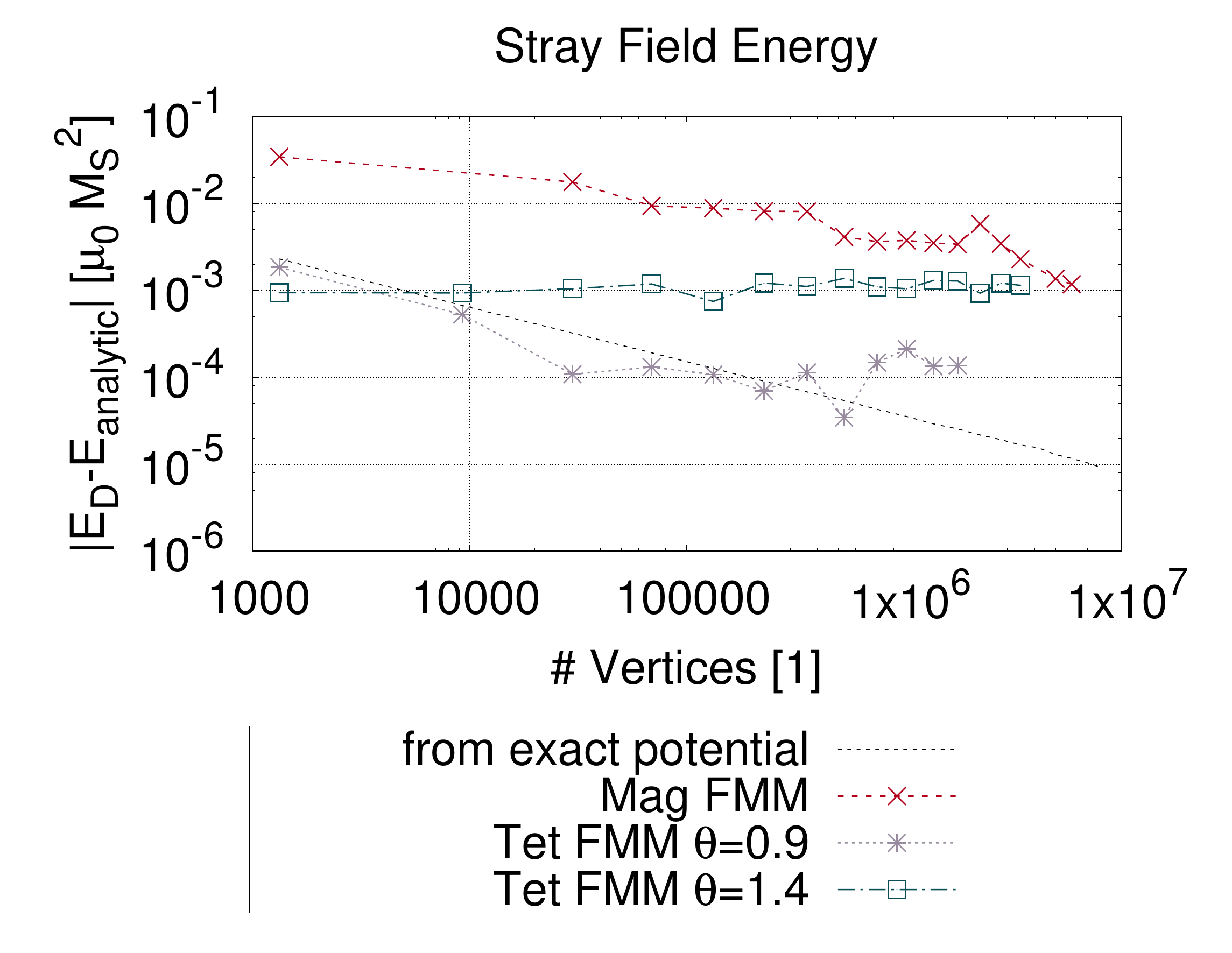}
\caption{This plot shows the stray-field energy error for the algorithms compared in \cref{fig:solvetime}. }
\label{fig:accuracy}
\end{figure}

\begin{figure}
\stdgrafx{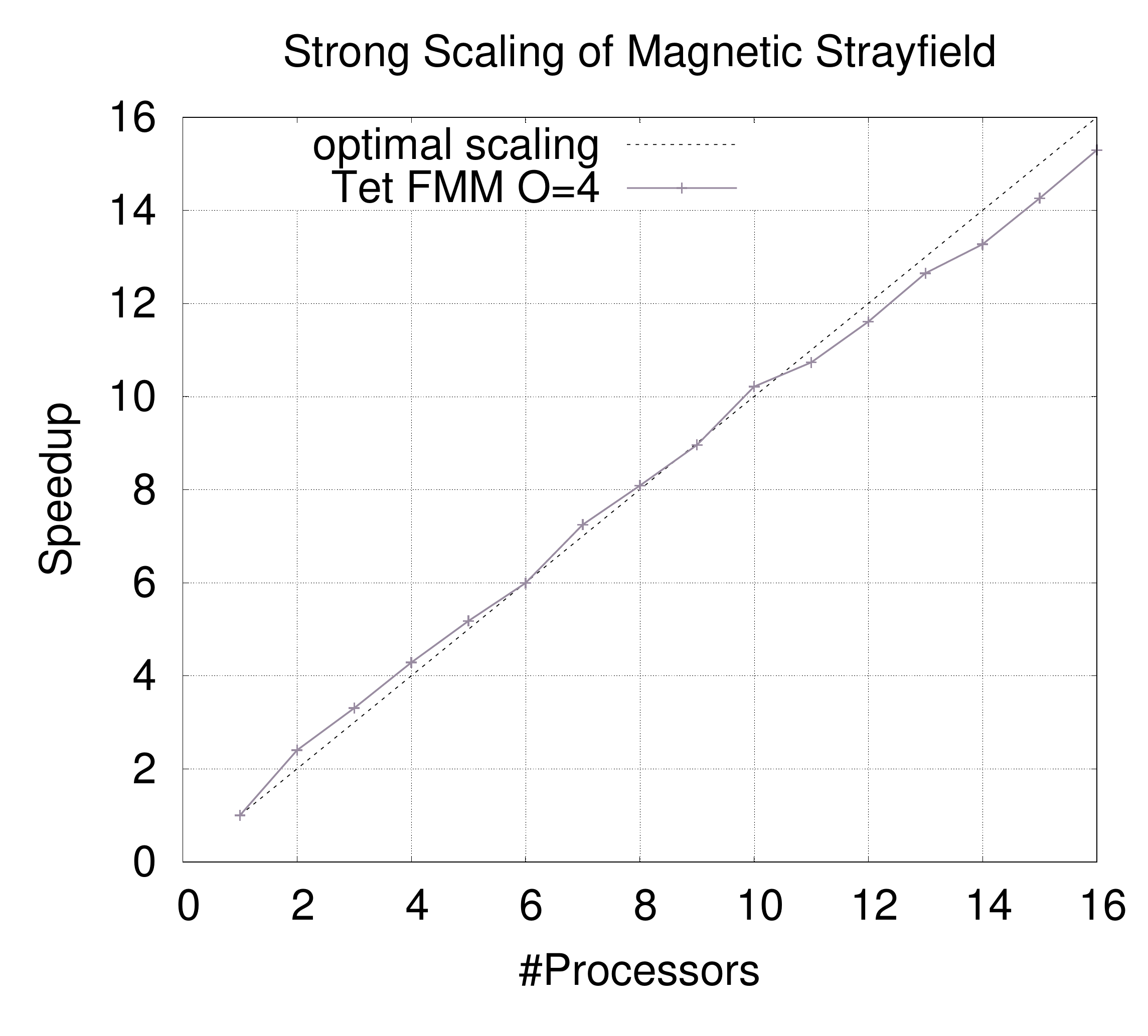}
\caption{This plot shows how the TetFMM code scales on a single node up to 16
  cores. Simulations were done with a $\theta=0.3$ and a multipole order $O=4$
  on \numprint{357911} vertices.}
\label{fig:scaling}
\end{figure}

\Cref{fig:scaling} shows a performance comparison of the tested methods. The Tet
FMM simulations start off a little slower because distributed computation is not
efficient for the
small meshes. With that
in mind, the Tet FMM codes show better scaling in the compute intensive region. The fourth line (Mag
FMM) shows the upper performance limit for the FMM using a $\theta=0.3$ with the
fastest current implementation using point-like dipole sources.
For a similar accuracy (\cref{fig:accuracy}) the FMM code is faster by a factor
of $\approx 3$ for \numprint{10000} particles and providing the
option of even faster evaluation if the necessary accuracy is smaller or more
cores are available. A small performance improvement should be possible by
varying maximum multipole order $O$ and MAC parameter $\theta$ for optimal
performance.

\Cref{fig:accuracy} shows the difference of the stray-field energy to the
analytically computed energy ($E_{analytic}=1/6 \mu_0 M_S^2$). The E error is
not constant because the multipole levels increase with the problem size and the
real $theta$ depends on the varying octree dissection of the mesh. As an
estimate the constant error for more than 1 million vertices results to $\Delta
E \approx 10^{-3} \mu_0 M_S^2$ for maximum multipole order $O=4$ and a $\Delta E
\approx 3\times 10^{-4} \mu_0 M_S^2$ for $O=6$ respectively. A smaller error can
be achieved by changing the MAC $\theta$ and multipole order at a performance cost.

The error of the Mag FMM code is caused by the non-continuous nature of
the sources; it is much lower for atomistic scales when sources are set
into the crystal lattice points and the problem becomes non-continuous. The
theoretical optimum was computed using the analytically computed potential at
each vertex. If the $H$-field is computed directly --- which is the case for
some other methods --- lower errors can be achieved.

To show the scaling properties of the code strong scaling on a single node is
shown in \cref{fig:scaling}. It is easy to see that the code scales nearly
optimally up to the 16 available cores. This indicates that it is compute-
instead of memory-bound and that it would profit more by the next hardware
generations than most other methods currently in use.

\section{Conclusion}
Nowadays, many core algorithms' bottleneck is transfer speed and the trend seems
to continue in that direction. The scaling properties of FMM are linear up to
many processors and even GPUs\cite{yokota_tuned_2012}. The numeric results show that FMM computation
today is comparable in speed to fast FEM implementations when the potential is
required for calculation. In short the presented FMM is a good fit for
micromagnetic calculations now and in the future.

\section{Acknowledgment}
The authors acknowledge the Vienna Science and Technology Fund (WWTF) under Grant No. MA14-044, the Austrian Science Fund (FWF): F41 SFB ViCoM and the the CD-Laboratory AMSEN (financed by the Austrian Federal Ministry of Economy, Family and Youth, the National Foundation for Research, Technology and Development) for financial support. The computational results presented have been achieved using the Vienna Scientific Cluster (VSC).

\appendix
\crefalias{section}{appsec}
\section{Explicit Integrals for the Direct Integration} \label{app:directIntegrals}
\begin{equation}{\int_T\frac{\mathbf{1}}{{\left( x_t^2+y_t^2+h^2 \right)}^{\frac{3}{2}}} \dd x_t \dd y_t}
= \int_{y-}^{y+} \left(\frac{x}{(h^2+y^2)\sqrt{h^2+x^2+y^2}} \right)\Bigg|_{x=x-}^{x+}\dd y_t \end{equation}
\rule{0pt}{4ex}
\begin{equation}
{\int_T\frac{\mathbf{y_t^2}}{{\left( x_t^2+y_t^2+h^2 \right)}^{\frac{3}{2}}}\dd x_t \dd y_t} \\
= \int_{y-}^{y+} \left(\frac{x y^2}{\left(h^2+y^2\right) \sqrt{h^2+x^2+y^2}}\right)\Bigg|_{x=x-}^{x+}\dd y_t
\end{equation}
\rule{0pt}{4ex}
\begin{equation}\begin{aligned}&{\int_T\frac{\mathbf{x_t}}{{\left( x_t^2+y_t^2+h^2 \right)}^{\frac{3}{2}}}\dd x_t \dd y_t}
= -\int_{y-}^{y+} \left(\frac{1}{\sqrt{h^2+x^2+y^2}}\right)\Bigg|_{x=x-}^{x+} \dd y_t \\
&= \frac{1}{\sqrt{h^2+\frac{{(x_1 (y_2-y)+x_2 (y-y_1))}^2}{{(y_1-y_2)}^2}+y^2}}
  -\frac{1}{\sqrt{h^2+\frac{{(x_1 (y_2-y)+x_3 (y-y_1))}^2}{{(y_1-y_2)}^2}+y^2}}
\end{aligned} \end{equation}
\rule{0pt}{4ex}
\begin{equation} \begin{aligned}
& {\int_T\frac{\mathbf{x_t y_t}}{{\left( x_t^2+y_t^2+h^2 \right)}^{\frac{3}{2}}} \dd x_t \dd y_t}
= -\int_{y-}^{y+} \left(\frac{y}{\sqrt{h^2+x^2+y^2}} \right)\Bigg|_{x=x-}^{x+}\dd y_t \\ 
& = f_{xy}(b1,k1,y)+g_{xy}(b1,k1,y)-f_{xy}(b2,k2,y)+g_{xy}(b2,k2,y)
\end{aligned} \end{equation}
with
$$f_{xy}(b,k,y) = \frac{\sqrt{b^2+h^2+2bky+y^2(1+k^2)}}{1+k^2}$$
$$g_{xy}(b,k,y)=\frac{b k \ln(b k + y+k^2y+\sqrt{1+k^2}\sqrt{b^2+h^2+2bky+(1+k^2)y^2})}{{(1+k^2)}^\frac{3}{2}}$$
$$b1=x_2-y_2 k_1;\;k1=\frac{x_1-x_2}{y_1-y_2}$$
$$b1=x_3-y_2 k_3;\;k2=\frac{x_1-x_3}{y_1-y_2}$$
\rule{0pt}{4ex}
\begin{equation}\begin{aligned}
&{\int_T\frac{\mathbf{y_t}}{{\left( x_t^2+y_t^2+h^2 \right)}^{\frac{3}{2}}}\dd x_t \dd y_t} \\
& = \int_{x_{y-1}}^{x_{y+1}}{\left(\frac{1}{\sqrt{h^2+x^2+y^2}}\right)}\Bigg|_{y=y_{y-}}^{y=y_{y+}} \dd x_t \\ 
& + \int_{x_{y-2}}^{x_{y+2}}{\left(\frac{1}{\sqrt{h^2+x^2+y^2}}\right)}\Bigg|_{y=y_{y-}}^{y=y_{y+}} \dd x_t \\
& = -f_2(b1y,k1y,y_1)-f_2(b2y,k2y,y_2)+2\ln\left(x+\sqrt{h^2+x^2+y_2^2}\right)
\end{aligned}\end{equation}
$$f_2=\frac{\ln(bk+x+k^2x+\sqrt{1+k^2}\sqrt{b^2+h^2+2bkx+x^2+k^2x^2})}{\sqrt{1+x^2}}$$
$$b1y=y_2-x_2 k1y;\; k1y=\frac{y_1-y_2}{ x_1 - x_2}$$
$$b2y=y_1-x_1 k2y;\; k1y=\frac{y_1-y_2}{ x_1 - x_3}$$

\bibliography{all}
\end{document}